\documentclass[prl,aps,showpacs,twocolumn,floatfix, superscriptaddress]{revtex4-1}
\usepackage{graphicx}
\usepackage[ansinew]{inputenc}
\usepackage{array}
\usepackage{color}
\usepackage{amsmath}
\usepackage{amsxtra}
\usepackage{amstext}
\usepackage{amssymb}
\usepackage{latexsym}
\usepackage{dsfont}

\begin{document}

\title{Dissipation-driven nonclassical state generation in optomechanics with squeezed light}

\author{Jae Hoon Lee}
\affiliation{Korea Research Institute of Standards and Science, Daejeon 34113, South Korea}
\author{Junho Suh}
\affiliation{Korea Research Institute of Standards and Science, Daejeon 34113, South Korea}
\author{Hyojun Seok}
\email{hseok@kongju.ac.kr}
\affiliation{Department of Physics Education, Kongju National University, Gongju 32588, South Korea}

\begin{abstract}

	We study an optomechanical system for the purpose of generating a nonclassical mechanical state when a mechanical oscillator is  quadratically coupled to a single-mode cavity field driven by a squeezed optical field. The system corresponds to a regime where the optical dissipation dominates both the mechanical damping and the optomechanical coupling. We identify that multi-phonon processes emerge in the optomechanical system and show that a mechanical oscillator prepared in the ground state will evolve into an amplitude-squared squeezed vacuum state. The Wigner distribution of the steady state of the mechanical oscillator is non-Gaussian exhibiting quantum interference and four-fold symmetry. This nonclassical mechanical state, generated via reservoir engineering, can be used for quantum correlation measurements of the position and momentum of the mechanics below the standard quantum limit.
\end{abstract}

\maketitle
Advancements towards ground state cooling of mechanical systems~\cite{Cooling1, Cooling2} have proven to deliver major breakthroughs for new developments in the quantum control of massive objects' mechanical motion. These developments now facilitate intensive exploration of cavity optomechanics in the deep quantum regime~\cite{Review1, Review2, Review3}. Important achievements particularly include the demonstration of macroscopic quantum states such as the motional ground state~\cite{Ground_state1, Ground_state2}, the conditional single-phonon state~\cite{Conditional_single_phonon_state}, and the mechanical squeezed state below zero-point fluctuations~\cite{Mechanical_squeezed_state}. Up-to-date, most optomechanical systems operate in the so-called weak-coupling regime where the optomechanical interaction can be described by a beam-splitter type interaction between photons and phonons, confining the mechanics to Gaussian steady states~\cite{Review1, Review2, Review3}. In the single-photon strong coupling regime, it has been proposed that the intrinsic nonlinearity of the optomechanical interaction can manifest itself giving rise to non-Gaussian~\cite{nonGaussian}, and nonclassical antibunched mechanical states~\cite{Antibunched_phonon1, Antibunched_phonon2, Antibunched_phonon3, Antibunched_phonon4}.

Much theoretical and experimental research in quantum optics focus on the employment of squeezed light~\cite{Squeezed_light}, where the quantum fluctuations in one of the quadratures are less than the vacuum fluctuations, to the optical control of a mechanical motion. In this case, the phase-sensitive nature of squeezed light assist various applications including efficient laser cooling of atoms or trapped ions~\cite{Cooling_via_squeezed_light1, Cooling_via_squeezed_light2, Cooling_via_squeezed_light3},  amplified optomechanical coupling~\cite{Squeezed_optomechanics},
enhanced quantum synchronization~\cite{Squeezed_drive3}, and generation of mechanical squeezing~\cite{Squeezed_drive1}. As of late, it has been reported that squeezed light can suppress the radiation pressure noise~\cite{Squeezed_RPF} and enhance sideband cooling efficiencies of mechanical systems beyond the quantum back-action limit~\cite{Squeezed_drive4}.

In this work, we theoretically analyze a scheme to generate 
a nonclassical mechanical state in a quadratic optomechanical setup ~\cite{Middle1, Quadratic_coupling_photonic_crystal} via photonic reservoir engineering~\cite{reservoir_engineering} in the weak-coupling regime. We show that the optomechanical system undergoes four-phonon processes leading to the mechanical state being relaxed to a robust four-phonon coherent state. This is provided that the optical cavity field is driven by an external squeezed light and the photon dissipation is the prevailing source of damping in the composite system. 
The physics underlying the four-phonon process is that photons supplied by the squeezed reservoir are created or destroyed in pairs while a single photon is converted to two phonons in the quadratic optomechanical system. This results in a net four-phonon process for the mechanics, making the steady-state Wigner distribution a four-fold symmetric non-Gaussian distribution. Furthermore, we find that the steady-state Wigner distribution possesses negative regions indicating that the mechanical steady state is nonclassical. It is also shown that the mechanical steady-state displays squeezing in one of the squared quadratures of the mechanical motion. This attribute may be utilized to improve the precision in correlation measurements between position and momentum of the mechanical oscillator.

We consider an optomechanical system where the single-mode of a mechanical oscillator of mass $m$ and resonant frequency $\omega_m$ is quadratically coupled to a single-mode cavity field of frequency $\omega_c$ driven by an external monochromatic field of frequency $\omega_{L}$ at rate $\eta$. In the frame of the driving frequency for the cavity field, the Hamiltonian depicting the optomechanical system is $(\hbar =1)$
\begin{eqnarray}
 \hat H &=& -\Delta_c \hat a^\dag \hat a + i(\eta\hat a^\dag- \eta^*\hat a) + \omega_m \hat b^\dag\hat b\nonumber  \\ 
 &&+ g_0^{(2)}\hat a^\dag \hat a (\hat b+\hat b^\dag)^2 + \hat H_{\rm diss}. 
\end{eqnarray}
where $\Delta_c=\omega_L-\omega_c$ is the detuning from the cavity resonance, $g_0^{(2)}$ is the quadratic single-photon optomechanical coupling coefficient, and $\hat a$ ($\hat b$) is the annihilation operator for the optical (mechanical) field. Finally, $ \hat H_{\rm diss}$ accounts for dissipation and decoherence due to the coupling to the squeezed optical reservoir and mechanical bath at finite temperature.

If the external optical driving is strong enough that the steady-state mean intracavity photon number $n_c = \langle\hat a^\dag \hat a\rangle
 \gg 1$, the optomechanical interaction can be linearized with respect to the optical field operators. This linearization is expressed as $g_0^{(2)}\hat a^\dag \hat a (\hat b+\hat b^\dag)^2\rightarrow g_2(\hat a+\hat a^\dag)(\hat b+\hat b^\dag)^2$ where $g_2=g_0^{(2)}\sqrt{n_c}$ is the cavity-field amplified optomechanical coupling strength. Moreover, in this regime, the mechanical oscillator becomes stiffer as its resonance frequency is shifted as $\omega_m'=\omega_m+2g_0^{(2)}n_c$. As detailed in Ref.~\cite{Previous_paper}, if we assume that the external driving field is detuned by $\Delta_c=-2\omega_m'$ in the weak-coupling and resolved-sideband limit, we obtain, in the interaction picture with respect to the free field Hamiltonian, the following master equation for the density operator $\rho$ of the composite system:
\begin{equation}
\dot{\rho} = {\cal L}_{om}\rho+{\cal L}_o\rho+{\cal L}_m\rho,
\end{equation}
where ${\cal L}_{om}\rho=- ig_2[\hat a^\dag\hat b^2+\hat b^{\dag 2}\hat a,\rho]$ accounts for the coherent process in which phonons are created and destroyed in pairs accompanied by the annihilation and creation of one photon in the cavity field at rate $g_2$. Damping of the cavity field, at a rate $\kappa$, due to the squeezed optical reservoir is taken into account by
\begin{equation}
 {\cal L}_o\rho = \kappa{\cal D}[i\mu\hat a -i\nu\hat a^{\dag}]\rho,
\end{equation}
where the Lindblad superoperator is given by $ {\cal D}[\hat o]\rho = \hat o\rho\hat o^\dag-\hat o^\dag\hat o\rho/2 -\rho\hat o^\dag\hat o/2$. We assume that the optical reservoir is ideal such that the strength of the squeezing $r$ and the squeezing direction in phase space $\theta$ parameterize $\mu$ and $\nu$ as $\mu=\cosh r$ and $\nu=e^{i\theta} \sinh r$. Finally, ${\cal L}_m\rho=\gamma(n_{\rm th}+1){\cal D}[\hat b]\rho+\gamma n_{\rm th}{\cal D}[\hat b^\dag]\rho$ describes the mechanical field dissipation, at rate $\gamma$, due to the mechanical reservoir whose temperature is $T$ and the thermal occupation number is $n_{\rm th} = [\exp(\hbar\omega_m/k_B T)-1]^{-1}.$ 

In the regime where the optical damping dominates the mechanical dissipation and the optomechanical coupling, the cavity field tends to approach a squeezed state characterized by the optical reservoir on a timescale of $1/\kappa$. On timescales longer than $1/\kappa$, the density operator describing the composite system can be approximated as $\rho \approx \rho_m \otimes \rho_o$, where $ \rho_m$ is the reduced density operator for the mechanics and $\rho_o$ is the reduced density operator for the cavity field~\cite{Carmichael_book}. In this case, the relatively slower dynamics of the optomechanical system is governed by that of the mechanical field. In order to study the mechanical-field driven dynamics, we properly eliminate the reduced density operator for the cavity field by following the standard Born-Markov technique~\cite{Louisell_book, Carmichael_book}. 

The dynamics of the reduced density operator for the mechanics, in the scaled time $\tau = \gamma t$, is then described by the effective master equation 
\begin{equation}
 \frac{d{\rho_m}}{d\tau} =C_2{\cal D}[\hat J]\rho_m+(n_{\rm th}+1){\cal D}[\hat b]\rho_m+ n_{\rm th}{\cal D}[\hat b^\dag]\rho_m,
 \end{equation}
where $C_2= 4g_2^2/(\kappa\gamma)$ is the multiphoton optomechanical cooperativity for the quadratic coupling. Here, the effective coupling between the mechanical oscillator and the squeezed optical reservoir causes the mechanics to experience the {\it quantum jump} described by
\begin{equation}
 \hat J =\mu\hat b^2+\nu\hat b^{\dag 2}. \label{eq:jump}
\end{equation}
Note that the mechanical oscillator in this expression is coupled to two independent reservoirs: the intrinsic mechanical reservoir and an effective reservoir stemming from the squeezed optical reservoir. Naturally, the steady-state of the mechanical oscillator depends not only on the independent characteristics of the optical and mechanical reservoirs, but also on how the reservoirs are coupled to the mechanical oscillator. For the system of interest, where the coupling strength to the optical reservoir is much larger than that to the mechanical reservoir, the initial mechanical state quickly approaches a transient state directed by the squeezed optical reservoir. The state is then gradually disturbed by the mechanical reservoir and finally relaxes to an ultimate steady-state determined by couplings to both reservoirs. Provided that
\begin{equation}
 C_2|\nu|^2 \gg {n_{\rm th}}, 
\end{equation}
one can neglect the decoherence and dissipation due to the intrinsic mechanical reservoir under timescales when one phonon-losses are negligible $(\tau \ll 1)$. Then the dynamics of the mechanical oscillator can be reduced to
\begin{equation}
 \frac{d{\rho_m}}{d\tau} =C_2{\cal D}[\hat J]\rho_m.
\end{equation}

It is noteworthy to see that the mechanical oscillator experiences four-phonon processes in a single quantum jump. The four-phonon process can be readily interpreted as the following. First, the squeezed photons driving the optomechanics are supplied to the cavity in pairs. Since the optomechanical interaction is assumed to be quadratic, one of the photons assists a two-phonon absorption while its pair collaborates a parallel two-phonon emission, resulting in a net four-phonon process for the mechanics. It should be noted that the quantum jump operator $\hat J$ can be compared to the Bogoliubov-mode annihilation operator  
\begin{equation}
\hat \beta =\mu \hat b+\nu \hat b^{\dag}
\end{equation}
whose vacuum is the single-mode squeezed vacuum~\cite{Squeezed_vacuum}. In Ref.~\cite{Dissipative_squeezing1, Dissipative_squeezing2, Mechanical_squeezed_state}, a precooled mechanical oscillator is shown to relax to a robust squeezed state, experiencing a quantum jump described by the Bogoliubov-mode annihilation operator in a linear optomechanical setup with modulated coherent driving. However, in our scheme, quadratic coupling aided by the squeezed optical reservoir makes the mechanics go through higher-order quantum jumps and thus leads to higher-order squeezing.

\begin{figure}[]
\includegraphics[width=0.48\textwidth]{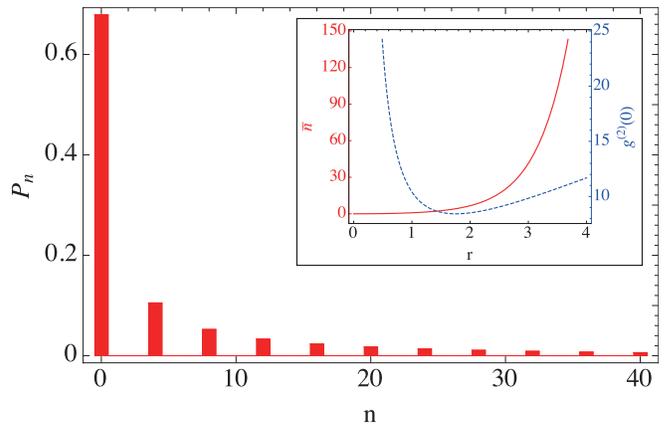}
\caption{
Steady-state phonon number distribution $P_n$ of the mechanical oscillator for the optical squeezing parameter $r=2$. Inset: Steady-state mean phonon number $\bar n$ (red solid line) and normalized second-order correlation function $g^{(2)}(0)$ (blue dashed line) of the mechanical oscillator as a function of the squeeze parameter.
}
\label{fig:number}
\end{figure}

The steady-state for the mechanical oscillator is required to satisfy $\hat J\rho =0$ and $\rho\hat J^{\dagger} = 0$. We find that if the mechanical oscillator is prepared in the ground state it evolves into a statistical mixture of states before being restored to a pure state. This type of behavior can also be seen in quantum optics where light is coupled to a two-photon absorber~\cite{Knight1, Knight2}.  Therefore, the density operator describing the steady-state for the mechanics can be written in terms of the eigenstate of the jump operator $\hat J$ with the zero eigenvalue as $\rho_{\rm ss}= | \psi\rangle\langle \psi |$, where $|\psi\rangle$ is the state vector such that 
\begin{equation}
 \hat J | \psi\rangle =0 \label{eq:jump_ground}. 
 \end{equation} 
Expanding $ |\psi\rangle$ in the Fock states basis as  $|\psi\rangle=\sum_{n=0}^{\infty}c_n| n\rangle$ leads to the recursion relation:
\begin{eqnarray}
c_2&=&0,\,\,\,\,\,\,\,\,c_3=0, \nonumber \\
 c_{n} &=& -\sqrt{\frac{(n-2)(n-3)}{n(n-1)}}\frac{\nu}{\mu}\,c_{n-4},\,\,\,\,\, n\geq4. \label{eq:recursion}
\end{eqnarray}
Note from the recursion relation that there are two distinct solutions, one involving only $|4m\rangle$ phonon states and a second involving only $|4m+1\rangle$ phonon states. It is clear from Eq.~(\ref{eq:recursion}) that a mechanical oscillator prepared in the ground state will evolve into the steady-state
\begin{equation}
| \psi\rangle= \frac{1}{\cal N}\sum_{m=0}^{\infty}\sqrt{\frac{\left(\frac{1}{2}\right)_m\left(\frac{1}{4}\right)_m}{\left(\frac{3}{4}\right)_m m!}}\left(-\frac{\nu}{\mu}\right)^m| 4m\rangle,
\end{equation}
where $(a)_m=\frac{\Gamma(a+m)}{\Gamma(a)}$ is the raising Pochhammer symbol, ${\cal N}=\sqrt{{}_2F_{1}(\frac{1}{2}, \frac{1}{4}; \frac{3}{4}; \frac{|\nu|^2}{\mu^2})}$ is the normalization constant, and ${}_2F_{1}(a, b; c; z)$ is the hypergeometric function. 
The phonon statistics of the mechanical steady-state reads
\begin{equation}
 P_{n}= \frac{1}{{}_2F_{1}(\frac{1}{2}, \frac{1}{4}; \frac{3}{4}; \frac{|\nu|^2}{\mu^2})}\frac{\left(\frac{1}{2}\right)_m\left(\frac{1}{4}\right)_m}{\left(\frac{3}{4}\right)_m m!}\left(\frac{|\nu|}{\mu}\right)^{2m}\delta_{4m, n} \label{eq:population}
\end{equation}
and is plotted in Fig.~\ref{fig:number}, highlighting that only every fourth phonon number state, including the ground state, is populated. In the regime where the optical squeezing is weak, $r\ll 1$, the jump operator becomes $ \hat J \rightarrow \hat b^2$ so that the initial ground state becomes a dark state from the two-photon absorption process and thus only the ground state is significantly populated. As the squeezing parameter of the optical reservoir is increased, the steady-state phonon number distribution becomes oscillatory and has a long tale in the distribution, akin to the power-law distribution of the thermal distribution~\cite{Milburn_book1}. The phonon distribution shown here is can be compared to that of the squeezed vacuum state, where only even number states are populated, whereas here only every fourth number state is populated. 

The inset in Fig.~\ref{fig:number} shows the steady-state mean phonon number obtained from Eq.~(\ref{eq:population})
\begin{equation}
 \bar n \equiv \langle\hat b^\dag \hat b\rangle = \frac{{}_2F_{1}(\frac{3}{2}, \frac{5}{4}; \frac{7}{4}; \frac{|\nu|^2}{\mu^2})}{{}_2F_{1}(\frac{1}{2}, \frac{1}{4}; \frac{3}{4}; \frac{|\nu|^2}{\mu^2})}\frac{2|\nu|^2}{3\mu^2},
 \end{equation}
 and the normalized second-order correlation function $g^{(2)}(0)=\langle\hat b^{\dag2} \hat b^2\rangle/\langle\hat b^\dag \hat b\rangle^2$ of the mechanical oscillator in the steady state as a function of the squeezing parameter. This calculation indicates that the steady-state mean phonon number grows monotonically with the squeezing parameter and the mechanical field exhibits super-Poissonian statistics involving highly correlated pairs of phonons. 

It is known that if any of Klyshko's figures of merit $K_n=[(n+1)P_{n-1}P_{n+1}]/[nP_n^2]$ $(n=1, 2, \cdots)$ is smaller than unity, the state is nonclassical and can be used as a relatively simple measure of nonclassicality by means of counting experiments~\cite{Klyshko}. 
For the mechanical steady-state shown here, Klyshko's figures of merit are zero for every fourth phonon number state, implying that the mechanical steady-state is nonclassical. Recent experiments employing optical probes with single photon detectors have allowed high fidelity phonon
counting measurements giving promise for the detection of these types of nonclassical features in a nonlinear optomechanical resonator~\cite{Phonon_counting}. 

Note that this nonclassical state is introduced in the context of the generalized squeezing criterion~\cite{Zoller}: when two observables $\hat A$ and $\hat B$ obey the commutation relation $[\hat A, \hat B]=i\hat C$ and thus satisfy the uncertainty relation $ \Delta \hat A \Delta \hat B \geq \frac{1}{2}|\langle \hat C\rangle |$, one of the observables may satisfy the inequality relation $(\Delta \hat A)^2 \leq \frac{1}{2}|\langle \hat C\rangle |$. The notion of amplitude-squared squeezing was proposed where the real or imaginary component of the square of the optical field amplitude can be squeezed~\cite{Hillery, Hillery2}. In the present work, the steady-state for the mechanics $|\psi\rangle$ is a special case of the amplitude-squared squeezed vacuum~\cite{amp_squeezed3}.  

\begin{figure}[]
\includegraphics[width=0.48\textwidth]{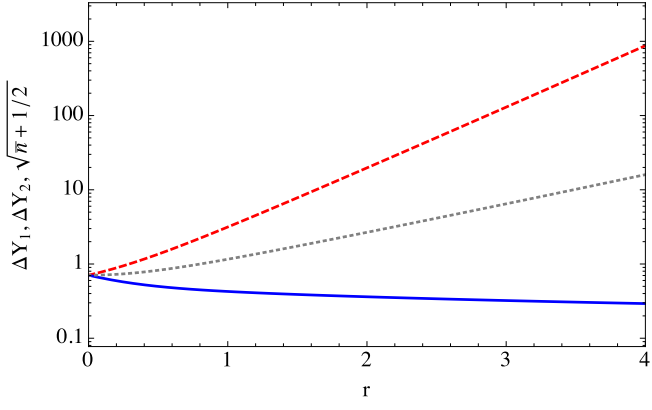}
\caption{
Quantum fluctuations in $Y_1$ and $Y_2$ as a function of the squeezing parameter $r$ along with the square-root of the lower bound of the uncertainty product $\bar n+1/2$, i.e., the standard quantum limit. The quantum fluctuations in $Y_1$ (blue solid line) is less than $\sqrt{\bar n+1/2}$ (gray dotted line) while the fluctuations in $Y_2$ (red dashed line) is larger than $\sqrt{\bar n+1/2}$. The product of the two fluctuations is equal to $\bar n+1/2$ regardless of the squeezing parameter.  
}
\label{fig:2}
\end{figure}

We now turn to investigate the phase-dependent properties of the steady-state mechanical oscillator. It is convenient to  introduce two hermitian operators describing the real and imaginary parts of the square of the mechanical mode as $ \hat Y_1 = \frac{1}{2}(\hat b^{2}e^{-i\theta/2} + \hat b^{\dag2} e^{i\theta/2})$ and $\hat Y_2 = \frac{1}{2i}(\hat b^2e^{-i\theta/2} - \hat b^{\dag2} e^{i\theta/2})$. Since the commutation relation between $\hat Y_1$ and $\hat Y_2$ is $[\hat Y_1, \hat Y_2] = i(2\hat b^\dag\hat b +1)$, the corresponding uncertainty relation is
\begin{equation}
 \Delta \hat Y_1 \Delta \hat Y_2 \geq \bar n+1/2.
\end{equation}
If we write down the jump operator $\hat J$ in terms of $\hat Y_1$ and $\hat Y_2$ in Eq.~(\ref{eq:jump_ground}) following the approach in Ref.~\cite{amp_squeezed3}, it is apparent that $\langle\hat Y_1\rangle$, $\langle\hat Y_2\rangle$, and $\langle\hat Y_1\hat Y_2+\hat Y_2\hat Y_1\rangle$ are all zeros and 
\begin{equation}
\Delta \hat Y_1 = e^{-r}\sqrt{\bar n+1/2},\,\,
\Delta \hat Y_2=e^{r}\sqrt{\bar n+1/2}. 
\end{equation}
from the fact that the state $|\psi\rangle$ satisfies $\langle\hat J^2\rangle=\langle\hat J^\dag\hat J\rangle=0$. 
Therefore, $|\psi\rangle$ is not only the minimum uncertainty state for the amplitude-squared variables $Y_1$ and $Y_2$, but it also exhibits squeezing along $Y_1$: the variance of $Y_1$ is always less than the bound of the uncertainty product. This behavior in is shown in Fig.~\ref{fig:2} where we plot the quantum fluctuations of $Y_1$ and $Y_2$ as a function of the squeezing parameter along side the square-root of the bound of the uncertainty product.

\begin{figure}[]
\includegraphics[width=0.48\textwidth]{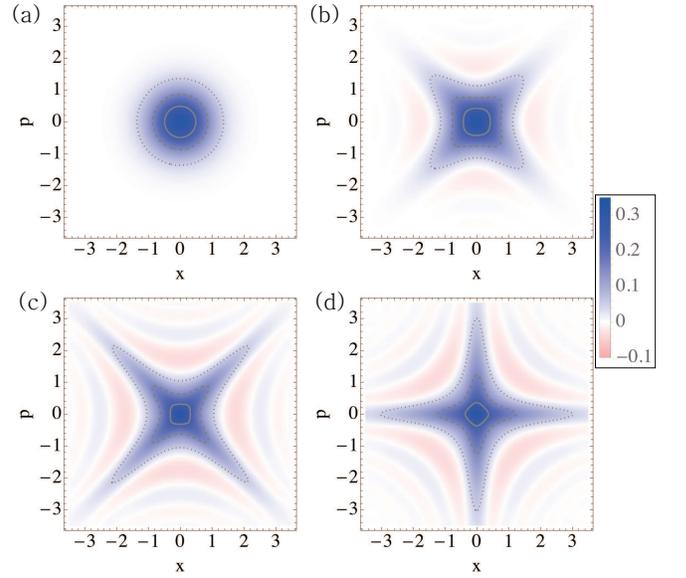}
\caption{Steady-state Wigner distribution of the mechanics as a function of the squeezing parameter of the optical reservoir (a) $r=0$, $\theta=0$, (b) $r=0.5$, $\theta=0$ (c) $r=1.0$, $\theta=0$ (d) $r=1.0$, $\theta=\pi$. Equidensity lines of $W=0.25$, $W=0.15$, and $W=0.05$ are presented with a solid, dashed, and dotted lines, respectively.
}
\label{fig:wigner}
\end{figure}

Further understanding and insight of the steady-state mechanical oscillator can be gained by investigating the corresponding phase-space distribution. Specifically, we employ the Wigner quasiprobability distribution, $ W(x, p) = \frac{1}{\pi}\int_{-\infty}^{\infty}dy~\langle x+y|\rho_{\rm ss}|x-y\rangle e^{-2ipy}$, where $x,~p$ are dimensionless position and momentum of the mechanics, measured in units of $x_0=\sqrt{\hbar/m\omega_m'}$ and $p_0=\sqrt{m\hbar\omega_m'}$. 
Fig.~\ref{fig:wigner} shows the steady-state Wigner distribution of the mechanics for various squeezing parameters.  As illustrated in Fig.~\ref{fig:wigner}-(b) and (c), when $\theta =0$ i.e. $\hat Y_1=\hat x^2-\hat p^2$, the Wigner density is expanded in the directions of $ Y_1=0$ (i.e. $p=\pm x$), and displays the quantum interference along the contours $Y_1\neq0$~\cite{contour}. As illustrated in Fig.~~\ref{fig:wigner}-(d), when $\theta = \pi$ i.e. $\hat Y_1=\hat x\hat p+\hat p\hat x$, the Wigner distribution is stretched along the lines corresponding to $ Y_1=0$ (i.e.~$x=p=0$). 
The four-fold symmetry originates from the fact that only every fourth number state can be populated~\cite{amp_squeezed3}. The Wigner density of the mechanical oscillator is positive and high along the lines of symmetry while possessing quantum interferences with negative values between the symmetry lines, which is a signature of a nonclassical state. Since quantum fluctuations in $Y_1$ are below the standard quantum limit, it may have a potential application for precision measurement of the quantum correlation between the position and momentum of the mechanics below the standard quantum limit. 

To conclude, we have analyzed a scheme in which the mechanical state of a quadratic optomechanical system can be engineered to exhibit a robust nonclassical state with a Wigner distribution containing negative regions. We find that a mechanical oscillator initially prepared in the ground state will go through four-phonon processes and relax to a robust pure state if the optomechanical setup is driven by squeezed light. The mechanical steady-state has significant non-zero population in every fourth number state, producing amplitude-squared squeezing and a four-fold symmetric non-Gaussian Wigner distribution. For sufficiently high optomechanical cooperativity, single-quadrature homodyne techniques can be employed for quantum state tomography of mechanical states to verify its nonclassicality~\cite{Tomography}.

\acknowledgments
H.~S. acknowledges the National Research Foundation of Korea (NRF) grant funded by the Korea government (MSIP) (Grant No.~2015R1C1A1A01052349). J.~L. acknowledges the R\&D Convergence Program of NST (National Research Council of Science and Technology) of the Republic of Korea (Grant No. CAP-15-08-KRISS). J.~S. acknowledges the Basic Science Research Program through the NRF funded by the Ministry of Science and ICT (Grants No. NRF-2016R1C1B2014713 and No. NRF-2016R1A5A1008184).

\end{document}